\documentstyle[12pt]{article}
\begin{document}
\pagestyle{empty}
\def\r{\rightarrow}
\def\be{\begin{equation}}
\def\ee{\end{equation}}
\def\ben{\begin{eqnarray}}
\def\een{\end{eqnarray}}
\baselineskip 21pt plus .1pt minus .1pt
\pagestyle{empty}
\noindent
\begin{flushright}
hep-ph/9905470\\
SINP/TNP/99-20\\
May 1999\\
\end{flushright}
\begin{center}
{\large\bf{Bi-maximal Neutrino  
Mixing With SO(3) 
Flavour Symmetry}} 
\end{center}
\vskip .25in
\begin{center}
Ambar Ghosal
\end{center}
\vskip .1in
\begin{center}
Saha Institute of Nuclear Physics,\\
Theory Division,
Block AF, Sector 1, Salt Lake,\\ 
Calcutta 700 064, India\\
\end{center}
\vskip 24pt
\noindent
We demonstrate that an $SU(2)_L\times U(1)_Y$ model with extended 
Higgs sector  gives rise to bi-maximal  neutrino mixing through the 
incorporation of SO(3) flavour symmetry and discrete symmetry.
The neutrino and the charged lepton  masses are generated due to higher 
dimensional terms. The hierarchical structures of neutrinos and 
charged leptons are obtained due to inclusion of SO(3) flavour 
symmetry and discrete symmetry.The model can accommodate  the vacuum 
oscillation solution of solar neutrino problem, through reasonable 
choice of model parameters along with the atmospheric neutrino experimental 
result. 
\vskip .75in 
\noindent
PACS No. 12.60 Fr., 
14.60 Pq., 13.40 Em.\\
E-mail: ambar@tnp.saha.ernet.in
\newpage
\pagestyle{plain}
\setcounter{page}{2}
\noindent
Evidence in favour of neutrino oscillation 
(as well as neutrino mass) has been 
provided by the Super-Kamiokande 
(SK) atmospheric 
neutrino experiment [1] through the measurement 
of magnitude and angular distribution of 
the $\nu_\mu$ flux produced in the atmosphere due to 
cosmic ray interactions. Observed depletion of 
$\nu_\mu$ flux in earth has been interpreted as the 
oscillation of $\nu_\mu$ to some other 
species of neutrino. In a two flavour neutrino 
oscillation scenario, oscillation between 
$\nu_\mu$ - $\nu_\tau$, the experimental result 
leds to maximal mixing between two species
$\rm{Sin^22\theta\sim 0.82}$ with a mass-squared 
difference 
$\Delta$$m^2_{atm}$$\sim$ 
$(5\times {10}^{-4}-6\times {10}^{-3})$ 
$\rm{eV^2}$.
The solar neutrino experimental results [2] 
are also in concordance with the interpretation of 
atmospheric neutrino experimental result and the data 
provide the following values as 
$\Delta$$m^2_{e\mu}$$\sim$ 
$(0.8 - 2)\times {10}^{-5}\rm{eV^2}$, 
$\rm{Sin^22\theta}$ 
$\sim$ 1 (Large angle MSW solution) or 
$\Delta$$m^2_{e\mu}$$\sim$ 
$(0.5 - 6)\times {10}^{-10}\rm{eV^2}$, 
$\rm{Sin^22\theta}$ 
$\sim$ 1 (vacuum oscillation solution). 
Furthermore, the CHOOZ experimental result [3] 
gives the value of $\Delta m^2_{eX}< {10}^{-3}$
$\rm{eV^2}$ or $\rm{Sin^22\theta}_{eX}< 0.2$. In order 
to reconcile with the solar and atmospheric neutrino 
experimental results, a distinct pattern of neutrino 
mixing emerges, namely, bi-maximal neutrino mixing [4], 
in which $\theta_{12}$=$\theta_{23}= 45^o$, and if, the 
CHOOZ experimental result is interpreted in terms of 
$\nu_e - \nu_\tau$ oscillation, 
then $\theta_{31}< 13^o$.
\vskip 24pt
In the prsent work, we demonstrate that 
an $\rm{SU(2)_L\times 
U(1)_Y}$ model wih extended 
Higgs sector coupled with an 
SO(3) flavour symmetry [5,6] and discrete 
$\rm{Z_3\times Z_3^\prime\times Z_4}$ symmetry, 
gives rise to 
nearly bi-maximal neutrino 
mixing along with the vanishing 
value of 
$\theta_{31}$. 
We have also discussed the situation when 
the mixing is exactly bi-maximal.
Instead of three almost 
degenerate 
neutrinos [5,7], we obtain a 
hierarchical pattern of neutrino 
masses. The 
charged lepton masses are 
also hierarchical in nature 
  and this is due to the inclusion of 
SO(3) flavour symmetry, 
which, when spontaneously broken, 
gives rise to the desired hierarchy in mass [6].
The discrete 
$\rm{Z_3\times Z_3^\prime\times Z_4}$ 
symmetry 
prohibits unwanted mass terms in the charged lepton 
and neutrino mass matrices as well as when accompanied 
with the choice of residual phases of the mass matrices 
give rise to required mixing pattern. 
We consider soft 
discrete symmetry breaking terms in 
the scalar potential, 
which are also responsible to obtain 
non-zero values of the 
VEV's of the Higgs fields upon 
minimization of the scalar 
potential. The Majorana neutrino 
masses are obtained 
due to explicit breaking of 
lepton number through higher 
dimensional terms. The leptonic fields 
($\rm{l_{iL}, E_{iR}}$, 
i = 1, 2, 3 is the generation index) 
and the Higgs fields 
$(\chi, \xi, \phi_1, \phi_2, 
\phi_3, \xi_e, \xi_\mu,  \rm{h})$   
have the following representation contents :
\ben
\rm{l_{iL}}\;(1, 2, -1),\;\; 
\rm{E_{R}}\;(3, 1, -2),\;\;  
 \chi\;(3, 1, 0),\;\;  \xi\;(3,1,0),\nonumber\\
\phi_1\;(3, 1, 0),\;\; \phi_2\;(3, 1, 0),\;\;
 \phi_3\;(3, 1, 0),\;\;  h\;(1, 2, 1),\nonumber\\
\xi_e\;(1, 1, 0),\,\,  \xi_\mu\;(1, 1, 0)
\een
where the digits in the parentheses   
represent SO(3), 
$\rm{SU(2)_L}$ and $\rm{U(1)_Y}$ 
quantum numbers. 
The subscript of the $\phi$ Higgs 
fields denote the direction of 
development of non-zero 
VEV, such as $<\phi_1>$ = $(v_1, 0, 0)$ etc. 
and the 
subscript below $\xi$ fields 
denote the respective 
coupling with the charged leptons.   
Regarding Higgs content and 
the symmetry breaking 
pattern, the present model is analogous 
to a supersymmetric 
model discussed in Ref.[6], 
where the Higgs scalars 
are replaced by 'flavon' chiral 
superfields. Regarding the 
representation content of leptonic fields, 
three lepton doublets 
($l_{1L}$, $l_{2L}$, $l_{3L}$)
are forming an SO(3) triplet while 
the right-handed charged leptons 
$(e, \mu, \tau)$ are singlet under 
SO(3) in Ref.[6] and this is just opposite 
to our case.
 We consider the 
following discrete 
$Z_3\times Z_3^\prime\times Z_4$ 
symmetry transformation of the 
lepton and Higgs fields: 
\begin{flushleft}
\underline{\rm 
${Z_3\times Z_3^\prime \times Z_4}$ Symmetry} 
\end{flushleft}
\ben
\rm{l_{1L}\r i\alpha\omega l_{1L},\;\; 
l_{2L}\r -il_{2L},\;\;
l_{3L}\r -il_{3L},\;\;
E_R\r i\alpha E_R}\nonumber\\
\chi
\r \alpha\omega\chi,\;\; \xi\r i\xi,\;\; 
{\rm{h\r h}},\;\; \phi_1\r \alpha\omega^2\phi_1,
\;\; \phi_2\r\alpha^2\omega\phi_2,\nonumber\\
\phi_3\r \alpha^2\phi_3,\;\; 
\xi_e\r \alpha^2\omega^2\xi_e,\;\;
\xi_\mu\r -\alpha^2\omega\xi_\mu
\een
where $\omega$ and $\alpha$ are the generators of 
$\rm{Z_3}$ and $\rm{Z_3^\prime}$ group, respectively. 
The most general lepton-Higgs Yukawa interaction in the 
present model generating 
Majorana neutrino masses is given by 
\be
L_Y^\nu = \rm{ 
\beta_1\frac{(l_{1L} l_{2L})(\chi\chi)hh}{M_f^3}
+ 
\beta_2\frac{(l_{1L} l_{3L})(\chi\chi)hh}{M_f^3}
+
\beta_3\frac{(l_{2L} l_{2L})(\xi\xi)hh}{M_f^3}
}$$
$$+\rm{\beta_4\frac{(l_{2L} l_{3L})(\xi\xi)hh}{M_f^3}
+
\beta_5\frac{(l_{3L} l_{3L})(\xi\xi)hh}{M_f^3}
}
\ee
and the Yukawa interaction which 
is responsible for generation 
of charged lepton masses is given by  
\be
L_Y^E = \rm{ 
\beta_6\frac{(e_R.\phi_1) l_{2L}h\xi_e^2}{M_f^3}
+
\beta_7\frac{(e_R.\phi_1) l_{3L}h\xi_e^2}{M_f^3}
+
\beta_8\frac{(e_R.\phi_2) l_{1L}h\xi_\mu}{M_f^2}
}$$
$$+\rm{\beta_9\frac{(e_R.\phi_3) l_{2L}h}{M_f}
+
\beta_{10}\frac{(e_R.\phi_3) l_{3L}h}{M_f}.
}\ee
In the above Lagrangian, 
we consider all the couplings 
$\beta_1$....$\beta_{10}$
are complex and are given by 
$\beta_i = |\beta_i|e^{i\delta_i}$
(i = 1,...10). For our analysis, 
we consider  
$|\beta_i|$ = 1. Among
ten phases, it is possible to absorb 
any five of them by redefining lepton fields 
and among the residual 
five phases, we set 
$\delta_2$ = $\delta_4$ = 
$\delta_7$ = $\pi$ and rest 
of them equal to 0. Although the dynamical origin of 
such choice of phases 
is not clear ( also not prohibited )
in the present framework, however, 
from the model building point of view such choice 
plays important role to obtain viable 
phenomenological scenario.  
 We  
 also consider
all the VEV's are real. 
The present model contains a large mass scale $M_f$, 
and for our analysis we set 
$\rm{M_f\sim M_{GUT}}$. We also 
consider that the 
flavour symmetry group SO(3) is 
broken below the GUT scale 
, but much above the electroweak 
scale corresponds to 
$<\rm{h}>$. On the otherway, 
the scale of SO(3) 
symmetry breaking 
VEV's, $<\chi>$ and $<\xi>$ 
are constrained by the solar 
and atmospheric neutrino 
experimental results, and the 
VEV's of $<\phi_1>$, $<\phi_2>$, 
$<\phi_3>$, $<\xi_e>$ and      
$<\xi_\mu>$ determine the masses 
of the charged leptons.  
The Higgs fields $\xi_e$, $\xi_\mu$ 
are singlet under the gauge 
symmetry and their VEV's in 
principle can take values above the 
SO(3) symmetry breaking scale.\hfill
\vskip 24pt
\noindent
In order to avoid any zero values of the 
VEV's of the 
Higgs fields upon minimization of the 
scalar potential, we have 
to consider discrete symmetry 
breaking terms. Without 
going into the details of the scalar 
potential, this feature 
can be realized in the following way. 
In general, the scalar 
potential can be written 
as (keeping upto dim=4 terms)
\be
\rm{
V = Ay^4 + By^3 + Cy^2 + Dy + E
}
\ee
where '$y$' is the VEV of 
any Higgs field and A, B, C, D, E are 
generic couplings of the terms 
contained in the scalar potential. 
Minimizing the scalar  
potential w.r.t. '$y$', we obtain
\be
\rm{
V^\prime = A^\prime y^3 + 
B^\prime y^2 + C^\prime y + D
}
\ee
Eqn.(6) reflects the fact 
that as long as $\rm{D}\neq 0$, and 
$A^\prime$ or $B^\prime$ or $C^\prime$ 
is not equal to zero, we 
will get non-zero solutions for '$y$'. 
Thus, in order to obtain $y\neq 0$ 
solution, it is necessary 
to retain the terms 
with generic 
coefficients D and  
$A^\prime$ or $B^\prime$ or $C^\prime$. 
In the present model, 
both the discrete symmetry breaking terms 
soft and hard, correspond to the term with 
coefficient D. Discarding hard symmetry breaking 
terms, we retain soft discrete symmetry breaking 
terms, and, hence, none of the VEV is zero upon 
minimization of the scalar potential.\hfill   
\vskip .1in    
\noindent
Let us look at the charged lepton 
sector. Substituting the 
VEV's of the Higgs fields appeared in Eqn.(4), we 
obtain the charged lepton mass matrix given by
\be
M_E = \pmatrix{0&d&-d\cr
               e&0&0\cr
               0&f&f}
\ee
where $d = \frac{<\phi_1><\rm{h}><\xi_e>^2}{M_f^3}$,  
$e = \frac{<\phi_2><\rm{h}><\xi_\mu>}{M_f^2}$ and   
$f = \frac{<\phi_3><\rm{h}>}{M_f}$. The hierarchy 
between the d, e and f parameters, 
$d<e<f$ is manifested 
due to the large mass scale $M_f$. 
Diagonalizing $M_{E}M_{E}^T $ 
, we obtain the following 
eigenvalues and mixing angles 
as 
\be
\rm{
m_{E_1} = \sqrt 2 d 
}$$
$$\rm{
m_{E_2} = e 
}$$
$$\rm{
m_{E_3} = {\sqrt 2} f
}
\ee
and $\theta_{12}^E$ = $\theta_{23}^E$ =   
$\theta_{31}^E$ = 0.   
It is to be mentioned that
the zero values of  
$\theta_{12}^E$, $\theta_{23}^E$ 
is assured due to discrete symmetry 
invariance 
however, the zero value of $\theta_{31}^E$ is 
obtained due to our choice of the value of 
$\delta_7$.
The hierarchy in the charged 
lepton masses  
arises due to the 
hierarchy already manifested in the mass matrix 
given in Eqn.(7). The three eigenvalues of  
the matrix $M_E$ can be 
fitted with the  
masses of the three 
charged leptons  
and the vanishing value $\theta_{31}^E$ is also  
not in conflict with the experimental value given by CHOOZ 
experiment as mentioned earlier.\hfill
\vskip .1in
\noindent
Let us now focus our attention to the 
neutrino sector of the model. 
Substituting the VEV's of $\chi$, $\xi$ and 
$\rm{h}$ Higgs fields in Eqn.(3), 
we get the Majorana 
neutrino mass matrix as follows: 
\be
M_\nu = \pmatrix{0&a&-a\cr
               a&b&-b\cr
               -a&-b&b}
\ee
where $a = \frac{<\chi>^2<\rm{h}>^2}{M_f^3}$,  
$b = \frac{<\xi>^2<\rm{h}>^2}{M_f^3}$. 
It is to be noted 
that the absence of $\nu_e\nu_e$ 
mass term in the above 
mass matrix (at the tree level) evades 
the bound on Majorana neutrino mass due to 
$\beta\beta_{o\nu}$ decay. 
Diagonalizing the neutrino 
mass matrix $M_\nu$ by an 
orthogonal transformation 
, we obtain the following values 
of the mixing angles, 
$\theta_{23}^\nu$ 
= $\frac{\pi}{4}$,  
$\theta_{31}^\nu$ = 0 and 
${\rm{tan^2}}\theta_{12}^\nu$ = 
$\frac{m_{\nu_1}}{m_{\nu2}}$. 
The 
eigenvalues of the above mass matrix comes out as
\be
-m_{\nu_1} = b - x 
$$
$$
m_{\nu_2} = b + x 
$$
$$
m_{\nu_3} = 0 
\ee
where $x$ = $\sqrt{b^2 + 2 a^2}$.  
It is to be noted that although the 
sign of $m_{\nu_1}$ can be made positive 
by setting appropriate values of 
residual phases, however, 
we will see that this can also be achieved 
due to phenomenological choices of model parameters. 
In the limit $b$$\r$ 0 ,$\theta_{12}$$\r$$\frac{\pi}{4}$,
the two eigenvalues $m_{\nu_1}$ and $m_{\nu_2}$ 
become degenerate and we can achieve the exact bi-maximal 
neutrino mixing. In this situation, 
although we obtain the exact bi-maximal neutrino 
mixing however, the obtained eigenvalues 
$m_{\nu_1}$ = $m_{\nu_2}$  and 
$m_{\nu_3}$ = 0, can be fitted with 
either the solar or the 
atmospheric neutrino experimental result. 
Removal of degeneracy between the two 
eigenvlues require further higher order  
corrections 
\footnote{
Almost degenerate neutrinos with 
bi-maximal mixing
can also be achieved by setting $\delta_2$ 
= 0 (instead of $\pi$) in Eqn.(3) and in this 
case we obtain 
$m_{\nu_1}$ = -${\sqrt 2} a$,
$m_{\nu_2}$ = ${\sqrt 2} a$ and 
$m_{\nu_3}$ = $2b$ with 
$\theta_{23}^\nu$ = 
$\theta_{12}^\nu$ = 
-$\frac{\pi}{4}$ and  
$\theta_{31}^\nu$ = 0.}. 
For oue analysis, we set the value  
of $\Delta m^2_{21}$ = $\Delta m^2_{sol}$ which 
in turn sets the value of $\theta_{12}$. 
The value of $x$ 
depends on the hierarchical relation between 
$a$ and $b$ parameters 
which is manifested from the values 
of $\Delta m^2_{21}$ = $4bx$ 
and $\Delta m^2_{23}$ = $(b+x)^2$. 
Now, if, $b^2>2a^2$, then 
the value of $x$ comes out 
as $x$  
$\sim b$ and, hence,   
$\rm{
m_{\nu_1} = m_{\nu_3}} = 0$ 
and $m_{\nu_2}$ = $2b$. In 
this situation also both the 
mass squared differences 
are parametrized in terms of a single 
parameter $b$, and, hence, in this 
case it is not possible 
to accommodate both the results of solar 
and atmospheric 
neutrino experiments. The same scenario appears 
for $a$ = $b$ case, and, hence, 
for a phenomenologically viable 
model, we have to consider 
the third option $2a^2>b^2$ and in this case 
$m_{\nu_1}$ is also become positive. 
In this situation, we obtain, 
$\Delta m^2_{21}$ = $4{\sqrt 2} ab$, 
$\Delta m^2_{23} = 
(a{\sqrt 2} + b)^2$ $\sim$ 
$2a^2$. For a typical value of 
$\Delta m_{23}^2$$\sim$ 
$4\times {10}^{-3}$ $\rm{eV^2}$  
which can explain the atmospheric 
neutrino deficits, 
we obtain 
$2a^2\sim 4\times {10}^{-3}$ $\rm{eV^2}$ 
which in turn gives rise to the value of 
$<\chi>$$\sim$ ${10}^{11}$ GeV 
for $M_f\sim {10}^{12}$ GeV 
and $<\rm{h}>\sim 100$ GeV. 
Using the same values of 
$M_f$ and $<h>$ parameters, 
we set  
the solar neutrino experimental 
result by setting the value of parameter 
$b$. 
For a typical value of 
$\Delta m^2_{21}$$\sim$$4\times {10}^{-10}$
$\rm{eV^2}$ which can explain 
the solar neutrino deficits due to  
vacuum oscillation, the value of $b^2$ 
comes out as $b^2\sim 0.25$ 
$\times {10}^{-17}$$\rm{eV^2}$ 
which leads to the value of $<\xi>$ 
$\sim$${10}^{7}$ GeV.  
The mixing angle $\theta_{12}^\nu$ in 
this case comes out as 
$\tan\theta_{12}^\nu$ = $\frac{a{\sqrt 2} - b}
{a{\sqrt 2} + b}$ and since 
$a>b$, $\theta_{12}^\nu\r$ $45^o$, and, hence,  
there is no conflict to satisfy the value of 
$\theta_{12}^\nu$ well within the allowed 
range of the experimental value. 
For 
large angle MSW solution, a typical 
value of  
$\Delta m^2_{21}$$\sim$${10}^{-5}$$\rm{eV^2}$ 
gives rise to 
$b^2\sim {10}^{-9}$$\rm{eV^2}$ 
and $<\xi>\sim$ $2\times {10}^9$ 
GeV, however, in this situation, since 
$b\sim a$ we obtain 
$\theta_{12}^{\nu}\sim \frac{\pi}{2}$, 
and , hence, large angle MSW solution is unlikely 
in the present model.\hfill
\vskip .1in
\noindent
In summary, we demonstrate 
that an $SU(2)_L\times U(1)_Y$ 
model with an extended 
Higgs sector, SO(3) flavour symmetry 
and discrete 
$Z_3\times Z_3^\prime\times Z_4$ 
symmetry, gives rise to nearly 
bi-maximal 
neutrino mixing   
$\theta_{23}$ = $\frac{\pi}{4}$, 
$\theta_{12}$$\sim$$\frac{\pi}{4}$ 
and 
$\theta_{31}$ = 0 
consistent 
with the present solar and 
atmospheric neutrino experimental 
results. Neutrino masses are 
generated due to explicit lepton 
number violating higher dimensional 
terms (dim=7) and the charged 
lepton masses are generated 
due to dim=5,6,7 terms. 
The hierarchical structure of charged 
lepton masses is obtained due to the 
inclusion of SO(3) 
flavour symmetry. The flavour diagonal 
structure of charged lepton mass matrix  
is also obtained due to our choice of 
residual phases in the charged lepton
mass matrix. Due to the same  
choice of the residual phases in the 
neutrino mass matrix, three non-degenerate 
neutrino masses are obtained with the value 
of the two mixing angles as  
$\theta_{23}$ = $\frac{\pi}{4}$ 
and $\theta_{31}$ = 0 respecting the atmospheic 
and CHOOZ experimental results, respectively.  
The value of $\theta_{12}$ depends on the masses of 
the neutrinos and in the exact limit of 
bi-maximal mixing 
($\theta_{12}$ = $\frac{\pi}{4}$), we obtain 
two degenerate neutrino masses. 
The discrete symmetry  
prohibits $\nu_e\nu_e$ mass term in the 
neutrino mass matrix at the tree 
level so as to evade the bound on the 
Majorana neutrino mass 
from $\beta\beta_{0\nu}$ decay in the present model. 
The vacuum neutrino 
oscillation solution 
can be achieved in the present model 
for a reasonable choice of model 
parameters along with the atmospheric 
neutrino experimental result 
whereas the large angle MSW solution is 
unlikely in the present model.\hfill
\vskip .25in   
Author acknowledges Yoshio Koide 
for many helpful 
suggestions and discussions.
\newpage
\begin{center}
{\large\bf{References}}
\end{center}
\begin{enumerate}
\item Super - Kamiokande Collaboration, 
 Y. Fukuda et al., Phys. Lett. B433, 
(1998) 9, {\it{ibid}} 436, (1998) 33, 
Kamiokande  Collaboration, 
S. Hatakeyama et al., Phys. Rev. Lett. 
81 
 (1998) 2016, T. Kajita, Talk presentd at 
 'Neutrino 98', Takayama, Japan, (1998). 

\item R. Davis, Prog. Part. Nucl. Phys. 32 (1994) 13; 
Y. Fukuda et al., Phys. Rev. Lett. 77 (1996) 1683, 
P. Anselmann et al., Phys. Lett. B357, (1995) 237, 
{\it{ibid}} B361, (1996) 235.
 
\item M. Appollonio et al. Phys. Lett. B420 (1998) 397.

\item V. Barger, S. Pakvasa, 
T. J. Weiller and  K. Whisnant, 
hep-ph/9806387, 
B. C. Allanach, hep-ph/9806294,
V. Barger, T. J. Weiller and K. Whisnant, 
hep-ph/9807319, J. Elwood, N. Irges and 
P. Ramond, Phys. Rev. Lett. 81, (1998) 5064, 
hep-ph/9807228, E. Ma, 
Phys. Lett. B442, (1998) 238,hep-ph/
9807386, hep-ph/9902392, 
G. Alterelli and F. Feruglio, 
Phys. Lett. B439 (1998) 112,hep-ph/9807353, 
Y. Nomura and T. Yanagida, Phys. Rev. 
D59 (1999) 017303, 
hep-ph/
9807325,  A. Joshipura, hep-ph/9808261, 
A. S. Joshipura and S. Rindani, 
hep-ph/9811252, 
K. Oda et al., Phys. Rev. D59, (1999) 055001, 
hep-ph/9808241, H. Fritzsch and 
Z. Xing, hep-ph/9808272, J. Ellis 
et al., hep-ph/
9808301, A. Joshipura and S. Vempati, hep-ph/
9808232, U. Sarkar, Phys. Rev. 
D59, (1999) 037302,
hep-ph/9808277,   
H. Georgi and S. Glashow , hep-ph/ 9808293;
A. Baltz, A. S. Goldhaber and M. Goldhaber, 
Phys. Rev. Lett. 81, (1998) 5730, 
hep-ph/ 9806540, 
M. Jezabek and A. Sumino, 
Phys. Lett. B440, (1998) 327, hep-ph/ 9807310,  
S. Davidson and S. F. King, 
Phys. Lett. B445 (1998) 191,
hep-ph/ 9808296,
 K. Kang, S. K. Kang, C. S. Kim and  
 S. M. Kim, 
 hep-ph/ 9808419, S. Mohanty, 
D. P. Roy and  U. Sarkar, 
Phys. Lett. B445, (1998) 185,
hep-ph/9808451;   
E. Ma, U. Sarkar and D. P. Roy, Phys. Lett. 
B444, (1998) 391, hep-ph/9810309,
B. Brahmachari, hep-ph/9808331,
R. N. Mohapatra and S. Nussinov, 
Phys. Lett B441, (1998) 299, 
hep-ph/ 9808301, Phys. Rev. 
D60, (1999) 013002, hep-ph/ 9809415,  
A. Ghosal, hep-ph/9903497, 
R. Barbieri, L.J.Hall and A. Strumia, 
Phys. Lett. B445 (1999) 407, 
hep-ph/9808333, Y. Grossmann, Y. Nir and 
Y. Shadmi, JHEP, 9810 (1998) 007, 
hep-ph/9808355, 
C. Jarlskog, M. Matsuda, S. Skadhauge and M. 
Tanimoto, 
Phys. Lett. B449, (1999) 240, 
hep-ph/9812282, S. M. Bilenky and 
C. Giunti, hep-ph/9802201, 
C. Giunti, Phys. Rev. D59, (1999) 077301,  
hep-ph/ 
9810272, 
M. Fukugita, M. Tanimoto and T. Yanagida, 
Phys. Rev. D57 (1998) 4429, 
S. K. Kang and C. S. Kim, 
Phys. Rev. D59, (1999) 091302,  
H. B. Benaoum and S. Nasri, hep-ph/9906232, 
C. H. Albright and S. M. Barr, hep-ph/9906297. 

\item D. O. Caldwell and R. N. Mohapatra, 
Phys. Rev. D50, (1994) 3477, A.S.Joshipura, 
Z.Phys. C 64 (1994) 31,
D. G. Lee and R. N. Mohapatra, Phys. Lett. B329, 
(1994) 463, P.Bamert and C.P.Burgess, Phys. Lett. B329, 
(1994) 289; E. Ma, hep-ph/9812344, 
Y. L. Wu, hep-ph/9810491, 
hep-ph/9901245, hep-ph/9901320, 
hep-ph/9905222. 

\item R. Barbieri, L. J. Hall, G. L. Kane 
and G. G. Ross, hep-ph/9901228, R. Barbieri, 
hep-ph/9901241.

\item A. S. Joshipura, Phys. Rev. D51(1995), 1321, 
A. Ioannissyan and J. W. F. Valle, Phys. Lett.  
B332, (1994) 93, A. Ghosal, Phys. Lett. B398, (1997) 315, 
A. K. Ray and S. Sarkar, Phys. Rev. D58, (1998) 055010. 

\end{enumerate}
\end{document}